\documentstyle[aps,pre,psfig,epsfig,multicol]{revtex}
\tightenlines
\newcommand{\npg}{\newpage}
\newcommand{\be}{\begin{equation}}
\newcommand{\ee}{\end{equation}}
\newtheorem{exercise}{Exercise}[section] 
\begin{document}
\title{\bf{Critical Phenomena: An Introduction  from a modern perspective
\footnote{Lectures given at the SERC School on ``Field theoretic methods in 
Condensed matter physics", held at MRI, Allahabad, India, Feb-Mar 2000} }}
\author{ Somendra M. Bhattacharjee} 
\address{Institute of Physics, Bhubaneswar 751 005, India}
\onecolumn
\date{\today}
\maketitle

Our aim in this set of lectures is to give an introduction to critical
phenomena that emphasizes the emergence of and the role played by
diverging length-scales.  It is now accepted that renormalization
group  gives the basic understanding of these phenomena and so, instead
of following the traditional historical trail, we try to develop the
subject in a way that emphasizes the length-scale based approach.

\begin{multicols}{2}
\tableofcontents
\end{multicols}
\npg
\section{Preamble}
\label{sec:preamble}

A phase transition is defined as a singularity in the free-energy or
any thermodynamic property of a system.  ``A system'' is to be
understood in a general sense; it is defined by a Hamiltonian under
various external macroscopic constraints.  For example, an isolated
collection of particles would be an example of a system, just as the
same collection exchanging heat or work with surrounding. It will also
be the same system even when it exchanges particles with surroundings.
The three examples given are the standard micro-canonical, canonical
and grand canonical ensembles in statistical mechanics. There are many
other types of ensembles. All these ensembles are expected to give the
same description of the system in the thermodynamic limit, at least
that's what says the principle of equivalence of ensembles.

Phase transitions are generally classified as first order, second
order ...etc, or first order and continuous type. 
 The behaviour of
any system near or at a continuous transition (also called critical
point) is distinctly different from the behaviour far away from it.
This peculiarity near a critical point is known as critical
phenomenon.

Power laws are distinctive features of critical phenomena and can be
ascribed to diverging length-scales.  This connection is now so
well-characterized that any phenomena, equilibrium or nonequilibrium,
thermal or nonthermal, showing power laws tend to be interpreted in
the same fashion of equilibrium critical phenomena through the
identification of relevant terms, scaling and diverging lengths.

In these notes, we shall concentrate on the general aspects of
critical phenomena. A few features of the first order transition will
be touched upon at some point.

One of the most important contributions of the studies of critical
phenomena is the shift in point of view to a length-scale based
analysis and classifying terms of a Hamiltonian (or in any other
description of a problem)  as {\it relevant} or
{\it irrelevant} rather than classifying them as numerically weak or strong.
The approach we take focuses on this and related issues.

Our approach is to show that the class of singular behaviour seen near
criticality can be characterized and understood via an emergent
diverging length scale as the transition point is approached.  This
way of characterizing singularities find straight-forward
generalization to many phenomena beyond equilibrium phase transitions.

Defining via singularity is the most general way of characterizing a
phase transition.  For a large class of systems, singularities could
occur due to ``ordering'' after a phase transition (or symmetry
breaking) but that is not necessarily a requirement for all
transitions. In other words, the existence of an ``order-parameter''
like quantity is not a prerequisite for understanding criticality
though it might be useful in many contexts.  We therefore deviate from
the conventional historical approach to the subject via mean field
theory, Landau theory, experimental observations leading to scaling.

Outline of the notes: The problem from a theoretical point of view is
presented, with a little bit of recapitulation of familiar things, in Sec
II to IV.  A possible resolution through scaling is introduced that
leads to finite-size scaling. The consequences and physical
interpretation leading to length-scale dependent parameters are
discussed in Sec. V and VI.  The Gaussian model and the $\phi^4$ theory
are used as examples in section VII.  The problems (exercises), even if not 
attempted,  should be read for continuity.
Only elementary calculus is used throughout, but the
reader is assumed to have the background of statistical mechanics at 
the level of Reif.

\subsection{Large system: Thermodynamic limit}
\label{sec:large-syst-therm}

In the canonical ensemble, the properties of the system with
Hamiltonian $H$ are obtained from the free energy,
\begin{equation}
\label{eq:1}
F = - k_B T \ \ln\  Z,
\end{equation}
where $k_B$ is the Boltzmann constant, $T$ the temperature and 
$Z = \sum e^{-\beta H}$ is the partition function with $\beta = 1/k_BT$.
If a phase transition is a
singularity of this free energy, then, for a simple 
well-defined $H$, it can occur only if $Z$ has zeros for real values of the
parameters. This cannot be the case for a finite-sized system because
$Z$ is a sum of a finite number of Boltzmann factors (which are
always positive).
The zeros of $Z$ or singularities of $F$ can only be for complex
values of the parameters.  
These complex zeros might have real limits when $N\rightarrow \infty$
( infinite number of terms to be added in $Z$) .  Then
and only then can a phase transition occur.
\begin{quote}
{\bf Conclusion:} A phase transition  cannot take place in a finite
system.  The thermodynamic limit,  $ N\rightarrow \infty$, has to be taken.
\end{quote}

\npg
\section{Where is the problem?}
\label{sec:where-is-problem}

Statistical mechanics starts with the idea of a micro-canonical
ensemble: the Boltzmann formula $S=k_B \ln \Omega$ where
$\Omega$ is the degeneracy or the number of states accessible to the
system under the given conditions of energy ($E$), volume ($V$),
number of particles ($N$), etc. 

Thermodynamics starts with the basic postulate of the 
existence of an `entropy' function $S$, so that 
\begin{equation}
\label{eq:2}
E = E(V,S,N), \quad {\rm or} \quad S=S(E,V,N) .
\end{equation}

One may  consider either a system of fixed number of
particles in equilibrium with surroundings or  a large
isolated system and focus on a small part which is in equilibrium with
the rest.  By releasing constraints (like $E=$const, or $V=$ const,  or
$N=$ const etc) one changes the ensemble. This is tantamount to
changing the thermodynamic potential required to describe the system.

Mathematically, the change of one thermodynamic potential to another
is achieved via Legendre transformation.  For example, from entropy to
temperature (micro-canonical to canonical) is done via
\begin{mathletters}
\begin{eqnarray}
\label{eq:7}
T &=& {\partial E(V,S,N)\over\partial  S}{\Big |}_{_{V,N}}, \\
\label{eq:7b}
F &=& E - TS, 
\end{eqnarray}
\end{mathletters}
where $F$ is now a function of $(V,T,N)$. This is possible if and only
if Eq. (\ref{eq:7}) can be inverted to eliminate $S$ on the rhs of 
Eq. \ref{eq:7b} in favour of $T$.  This inversion is {\it possible } if 
${\partial T}/{\partial S} \not = 0$. 
Similarly, we require $\partial p/\partial V \ \not = 0,$ 
$ \partial h/\partial M \ \not = 0$, or  $\partial\mu/\partial N \
\not = 0$ etc, where $\mu$ is the chemical potential, $h$  magnetic
field and $M$ magnetization (for a magnetic system), for change of
appropriate ensembles.

Since ${\partial S\over\partial T}$ is related to the  heat capacity of
the system, we see trouble if the specific heat {\it diverges} for
some values of the parameters of the system.  This looks like an
algebraic problem of the transformation but it is vividly reflected in the
argument for the equivalence of ensembles.  So let us
recollect that.

Take the case of canonical and micro-canonical ensembles. The two
ensembles are 
equivalent if the energy fluctuation in the canonical case is 
very small. First note that, if $F \propto N$, 
$\langle E\rangle$ and $\langle E^2\rangle - \langle E\rangle^2$ are
both proportional to $N$  (more on this
later).  The angular brackets indicate statistical mechanical
averaging.  A straight forward manipulation, 
using derivatives of the partition function, yields 
\begin{equation}
\label{eq:10}
C_V = \frac{1}{k_BT^2} \ (\langle E^2\rangle - \langle E\rangle^2) 
\quad \Longrightarrow \quad \frac{\Delta E}{E} =  k_BT^2
\frac{\sqrt{c_V}}{\sqrt N},
\end{equation}
where $\Delta E$ is the standard deviation of $E$ and $c_V$ is the
specific heat (heat capacity per particle). For a thermodynamic
system, $N\rightarrow \infty$, and this fluctuation goes
to zero, provided $c_v$ is finite. Hence the equivalence. \\
Well, the argument fails when $c_V\rightarrow \infty$.

By our definition of phase transitions, a diverging second
derivative of the free energy implies a phase transition. A
hypothetical possibility? No, it occurs in many systems in many 
different contexts.  Such points will be classified as critical
points.  Wait for a better definition of a critical point.

Critical points seem to be at odds with the conventional wisdom of
thermodynamics and statistical mechanics.

\npg
\section{ Recapitulation - A few formal Stuff }
\label{sec:few-formal-Stuff}

 A consistent thermodynamic description requires a few  postulates,  
some important ones of which are:\\
(i) Existence of ``entropy'' ( already mentioned)\\
(ii) Extensivity of $E$ and $S$.\\
(iii) Convexity of the free energy.\\
Postulate (ii) is a simple additivity property, while (iii) is required
for stability.  The formulation of  statistical mechanics guarantees
(iii) but not (ii).  We need to ensure (ii) to make contact with
thermodynamics and this restricts the form of the Hamiltonians  we
need to consider.  (See exercises III.6 and III.7 below).

\subsection{Extensivity}
\label{sec:Extensivity}

From our experience of macroscopic system, we expect that under a
rescaling $V\rightarrow bV, \  N\rightarrow bN,$ and $ S \rightarrow
bS$,  total energy should  change accordingly, i.e. 
\begin{equation}
\label{eq:3}
E(bV,bS,bN) = bE(V,S,N)\quad \quad({\rm homogeneous\  function}).
\end{equation}
By choosing $b = 1/N$, ($b$ is arbitrary), we have 
\begin{equation}
\label{eq:4}
E(V,S,N) = N \ {\cal E} (V/N, S/N),\quad {\rm where }\quad {\cal E}
(x,y) = E (x,y,1). 
\end{equation}

Note that ${\cal E}(v,s)$ is the energy per particle and being a
function of $v,s$ ( volume and entropy per particle), ${\cal E}(v,s)$
is independent of the overall size (e.g. $N$) of the system. This
proportionality to the number of particles (or volume) is {\it
  extensivity}.  

\begin{quote}
\begin{exercise}
 Additivity means $E(V_1,S_1,N_1) + E(V_2,S_2,N_2) =
  E(V_1+V_2, S_1+S_2, N_1+N_2)$.  Show that this implies
  Eq. \ref{eq:3}. 
\end{exercise}
\begin{exercise}
    Sanctity of extensivity: Remember Gibbs paradox and its
   resolution?
\end{exercise}
\end{quote}
 
The classical thermodynamics is based on Eq. \ref{eq:3} which, via the Euler
relation for homogeneous functions, gives
\begin{equation}
\label{eq:6}
S{\partial E\over\partial S} + V {\partial E\over \partial V} + N
{\partial E\over \partial N} = E = TS - pV + \mu N,
\end{equation}
defining $T = \partial E/\partial S, \ -p = \partial E/\partial V$ and
$\mu = \partial E/\partial N$, where the partial derivatives are taken
keeping the other variables constant. Each of the three new parameters
defined, $T,p$ and $\mu$, is a derivative of an extensive quantity
with respect to another extensive variable. Therefore $T,p,\mu$, and
variables like these are independent of the scale factor $b$ of
Eq.(3). Such quantities are called {\it intensive quantities}.
\footnote{Quantities like $v$, or $s$, an extensive quantity per unit
  volume or per particle, are also independent of the scale factor
  $b$.  To distinguish these from the other type, we may call them
  ``densities'' or ``fields'' (highly confusing for field theorists).
  In most cases, these could be distinguished from the context.}

\begin{quote}
\begin{exercise}
Take $b=1+\delta l$ and derive Eq. \ref{eq:6} for 
$\delta l \rightarrow 0$.
\end{exercise}
\begin{exercise}
 From Eq. (\ref{eq:3}) show that 
 \begin{equation}
   \label{eq:5}
F(bV,T,bN) = bF(V,T,N),\quad {\rm and} \quad F(V,T,N) = N\ f(v,T).
 \end{equation}
\end{exercise}
\end{quote}

\subsection{Convexity: Stability}
\label{sec:Convexity:-Stability}

We recognize that the derivatives needed for the Legendre transforms
(see Eqs.  ~(\ref{eq:7})) are derivatives of conjugate pairs: specific
heat, compressibility, susceptibility etc.  These derivatives (second
derivatives of free energy) are called {\it response functions},
because they measure the change in the extensive variable as the
externally imposed conjugate intensive quantity is changed.

A generalization of the derivation of Eq. \ref{eq:10} for a pair
$(y,\Phi)$ ($\Phi$ being the extensive variable and $y$ the conjugate
intensive variable) leads to
\begin{equation} 
\label{eq:11}
{\partial \langle \Phi\rangle\over\partial y} {\Bigg|}_{y=0} = {1\over k_BT} (
\langle \Phi^2\rangle -
\langle \Phi\rangle ^2 ){\Bigg |}_{y=0}  
\end{equation}
which connects the response function for $\Phi$ to the latter's variance.  The
latter being a positive definite quantity ensures that the response
function is of a particular sign only. This in turn shows that the
free energy as a function of $y$ can have only one type of curvature
(positive or negative) - a property known as convexity of the free
energy. This positivity of the response function guarantees
thermodynamic stability.  This is the third postulate of
thermodynamics mentioned above (quite often stated as the
maximization/minimization principle).  This connection between
response and fluctuation  plays a crucial role in the subsequent
development, especially in developing a correlation function based
approach.

The main points we need here are\\
(1) thermodynamic potentials are additive and therefore obey a
simple scaling.\\
(2) There are conjugate pairs of variables $(T,S)$, $(p,v)$,
$(\mu,N)$, etc where the first one in each set is intensive (i.e.
independent of the size) while the second one is extensive (i.e.
proportional to the size).  One may change variable (e.g., by
releasing a constraint) and this is the change of ensemble in
statistical mechanics.  Two bodies in contact in equilibrium need to
have equality of the relevant intensive quantities ({\sf remember
  zeroth law?}).

\subsubsection{Comments}
\label{sec:Comments-1}

\begin{multicols}{2}
\begin{itemize}
\item Equivalence of ensembles relies on the sharpness of the 
  probability distribution for say $E$.  The width of the distribution
  is related to the  corresponding response function.  For sharp
  distributions, the probability of $E$ for the combined system can be 
  taken as the product of the individual probabilities (i.e. 
$P_{V_1+V_2}(E_1+E_2)=P_{V_1}(E_1) P_{V_2}(E_2)$, whence follows
extensivity. Broad distributions would create problem here. 
\item Broad distributions imply large fluctuations.  It is therefore
  expected that fluctuations would play an important role in critical
  phenomena.  Fluctuations are  responsible for the problem
  with naive extensivity.  
\item Let us write $\Phi = \int d^d x \ \phi({\bf x})$, where
  $\phi({\bf x})$ is a local quantity (density).  We have written it
  as a $d$-dimensional integral, though it would be a sum for discrete
  systems.  The width of the probability distribution for $\Phi$
  around the mean $\langle \Phi\rangle$ is given by the corresponding
  "susceptibility" as given by Eq. (\ref{eq:11}).  Assuming
  translational invariance,
  $\langle\phi(x)\rangle=\langle\phi\rangle$, and denoting the
  response function per particle by $\chi$, the relative width of
  $P(\Phi)$ is given by $N^{-1/2} \sqrt{\chi}/\langle\phi\rangle$.
\end{itemize}
\end{multicols}

\begin{quote}
\begin{exercise}
A very important model that is used extensively is the Ising model.
\begin{equation}
\label{eq:39}
  H = -J\sum_{<ij>} s_i s_j - h \sum_i s_i, \quad  J,h>0,
\end{equation}
where the spins $s_i=\pm 1$ are situated on a $d$-dimensional lattice, the
interaction could be restricted to nearest-neighbours only (denoted by 
$<ij>$), and $h$ is an external magnetic field.\\
(1) Show that the free energy is extensive for any value of $T$ and
$h$, except possibly a particular point.  (Difficult).\\
  Show this extensivity at $T=0$ and $T=\infty$. (easy)\\
(2) How do we define the dimensionality of the lattice? (Hint: How does the
number of paths connecting two points grow with their separation?)
For uniform hyper-cubic lattices (linear, square, cubic,...) $d$ can
be uniquely defined by the number of nearest-neighbours.
\end{exercise}
\begin{exercise}
Consider now  more general Hamiltonians where each spin interacts with
every-other:
\begin{equation}
  \label{eq:37}
   H_{\rm bogus} = -J\sum_{i<j} s_i s_j, \quad {\rm and} \quad H_{\rm mf} =
   -\frac{J}{2N}\sum_{i,j} s_i s_j.\quad (N \ {\rm is\  total\
     number\  of\  spins}.)  
\end{equation}
Show that  there is a problem with extensivity for $H_{\rm bogus}$ but
not for $H_{\rm mf}$ (for $N\rightarrow\infty$).   Do this for $T=0$.
This model ($H_{\rm mf}$) gives the mean-field theory as an infinite
dimensional model.    $H_{\rm bogus}$ is to be dumped.
\end{exercise}
\begin{exercise}
Diamagnetism shows negative susceptibility.  Any problem with convexity?
\end{exercise}
\end{quote}
\npg

\section{Consequences of divergence - Problem with extensivity?}
\label{sec:cons-diverg}

Choose a quantity say the specific heat or susceptibility, that
diverges at the critical point $T = T_c$ in the thermodynamic limit.
For simplicity\footnote{We have deliberately chosen an intensive
  variable. The case of the conjugate extensive variable is left as a
  problem. (See Ex. V.5.)} we keep only $T$ as the control parameter
(all others being kept at their respective critical values).  Let us
denote this quantity  by $C(T,N)$ - a total quantity for the whole system -
obtained from the partition function of an $N$-particle system.

Extensivity requires that  $N^{-1} C(T,N)$ has an $N$-independent
limit for $N\rightarrow\infty$, say $c(T)$, so that for large
but finite $N$
\begin{equation}
\label{eq:12}
 C(T,N) = N\ c(T) + C_{\rm cor}(T,N).
\end{equation}
Here $C_{\rm cor}(T,N)$ is the correction term in an
asymptotic analysis.  An obvious expectation is 
${1\over N} C_{\rm cor}(T,N)\rightarrow 0$ as  $N\rightarrow\infty$.
Well, Eq. \ref{eq:12}  cannot be valid at $T=T_c$, because the
lhs is finite while $c(T_c) = \infty$ (by hypothesis), i.e. the
correction term has 
to be as large as the main term. This is not the only problem.   Right
at $T=T_c$, $N$ is the only 
parameter in hand, all others  at their respective critical values. 
If we want ${1\over N} C(T_c,N)$ to diverge as $N\rightarrow \infty$,
we need 
\begin{equation}
\label{eq:13}
 \frac{C(T_c,N)}{N} = C_0\, N^p, \quad (p>0), \Longrightarrow
\fbox{$ C(T_c,N)\sim N^{1+p}$},
\end{equation} 
which looks like a violation of Eq. (\ref{eq:3}).  We have assumed a
power law form in the above equation , but many other possibilities
exist.  Our choice is motivated by the fact that a large number of
systems (real or models) do show such power laws.

Extensivity, ad nauseum, is a consequence of additivity: small pieces
can be glued together to form a big piece with no change in property.
In a sense boundaries can be ignored.  This has to fail at the
critical point. But how?

\vbox{
\begin{figure}[htbp]
  \begin{center}
    \psfig{file=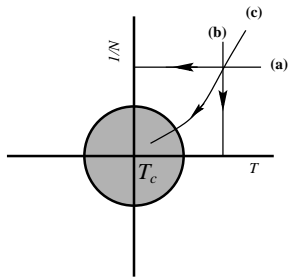,width=1.5in}
    \caption{Schematic diagram showing the various paths 
      in $T$ vs $N^{-1}$ plane, around the critical point
      $T=T_c$.} 
    \label{fig:1}
  \end{center}
\end{figure}
}

If we study the free energy, which cannot diverge, we might in
principle write down an expansion of the type proposed in
Eq. \ref{eq:12}.  This expansion is in $N$ for any $T$.  From such a 
free-energy, we might compute the specific heat by taking derivatives
and face again the problem of infinities on the right hand side at
$T=T_c$.  Another way of saying the same thing is that the free energy 
at the critical point behaves as $N^{1+q}$ with $q\leq 0$ but the
specific heat obtained by taking derivatives of the free energy with
respect to temperature gets a different power of $N$. This means that
$N$ and temperature occur as a combination variable and not
independently so that the limits $N\rightarrow\infty$ and 
$T\rightarrow T_c$ (taking derivatives)  may not commute. The double
limit needs to be considered carefully.  

The dilemma is shown schematically in Fig. 1.  The expansion in $N$,
Eq. (\ref{eq:12}), is valid along path (a) ( $T\rightarrow T_c$ for a
fixed $N<\infty$) and path (b) ( thermodynamic limit for a fixed $T$),
but the expansion is not valid in the neighbourhood of the critical
point.  The non-extensive behaviour gets reflected if the critical
point $(T=T_c, N^{-1}=0)$ is reached via a suitable path like
(c) whose form is obtained in the next section.  An isolated critical
point {\it does} affect its neighbourhood as in 
any problem of non-uniform convergence. The neighbouring region (shaded in
Fig \ref{fig:1}) is called critical region.

Question remains: Can the specific heat be divergent?

\subsubsection{Comments}
\label{sec:Comments-2}

\begin{multicols}{2}
\begin{itemize}
\item If the specific heat has the anomalous behaviour at $T=T_c$,
  then the energy-density is also expected to have so.  In general if
  the response function for a variable $<\Phi>$ has a singular
  behaviour, the variable  itself will also be singular.  In the next
  section we see how these are related.
\item With Eq. (\ref{eq:13})in hand, we need to redo the
  argument made in Sec II regarding fluctuation. This is done in
  Section ~\ref{sec:Exponents-relations}.
\item Question of divergence of specific heat: The answer is obviously
  ``no'' for a single particle system with a simple Hamiltonian
  (perfect gas, oscillator, two-level system ..).  For a
  noninteracting collection of such simple systems, the free energy is
  always proportional to $N$ and no criticality may occur even in the
  limit $N\rightarrow\infty$.  Interaction is needed and essential for
  criticality.

\end{itemize}
\end{multicols}
\npg

\section{ Generalized Scaling}

The simple scaling of Eq. \ref{eq:5} needs to be modified to allow for
the apparent non-extensive behaviour at a critical point. To show this
we work in the ensemble with $T$ as the variable (and all other
parameters are kept fixed at their critical values). More variables
will be taken up after that.   We start with a generalization of scaling
and show that it works. After working out a few consequences, we
discuss the physical significance of the original scaling hypothesis.

\subsection{One variable: temperature}
\label{sec:One-vari-temp}

 Let us define $t=(T-T_c)/T_c$.
Remember that this is an intensive variable and would not have
required any scale factor on rescaling as per Eq. \ref{eq:3} or Ex.
(\ref{eq:5}).  In order not to clutter the equations with extra
symbols, we consider a lattice problem for which $N$ and $V$ can be
interchanged (e.g. the Ising model of Eq.
~(\ref{eq:27})).  Defining $N=L^d$ with $L$ as a linear dimension of a 
$d$-dimensional cube, we
take a way out as
\begin{equation}
\label{eq:14}
F(t,L) = b^{-d+y_f} \ F(b^{-y_t} t, bL),\quad \quad({\rm generalized\
  homogeneity}), 
\end{equation}
from which  Eq. (\ref{eq:3}) or Eq. (\ref{eq:5})  can be
recovered for $y_f=y_t=0$.  We first show that this generalized
homogeneous function works and then, look at its consequences.
Physical significance is discussed after that.

Choosing $b = L^{-1}$,
\begin{equation}
\label{eq:15}
F(t,L) = L^d L^{-y_f} \ {\cal F}\left(\frac{t}{L^{-y_t}}\right ),
\Longrightarrow \ {\sf f}(t,L) = b^{-y_f} \ {\sf f}(b^{-y_t} t, bL),
\end{equation}
where ${\sf f}(t,L) \equiv {1\over N} F(t,L)$ is the free-energy per
unit volume.  We used ${\cal F}(x)=F(x,1)$. Finiteness of free-energy
requires $y_f>0$.

For $t \not =0$,  demand extensivity.  To get a linear dependence of $F$
on $N$, we require\footnote{ The sign $\sim$ is used to denote the
  functional dependence on the variables on the rhs, coefficients etc 
  may not be explicitly shown. Sign $\approx$ is to be used to denote
  the leading term or terms with all coefficients.} 
${\cal F}(x) \sim |x|^q$ such that $-y_f + q y_t = 0$. This gives 
\begin{equation}
\label{eq:16}
f(t) \equiv \lim_{L\rightarrow\infty} {\sf f}(t,L) \sim |t|^{y_f/y_t}, \quad 
{\rm or} \quad   f(t)=b^{-y_f} f(b^{y_t} t).
\end{equation}
Since the free energy per particle cannot be infinite even at $T_c$,
we need $y_f>0$ and $y_t >0$.  The last expression could also be
written as $f(t) = b^{-y_f/y_t} f(b t)$ by redefining $b$.

If we are right at the critical point, the free energy density behaves as
\begin{equation}
\label{eq:19}
f(t=0,L) \ \sim \ L^{-y_f},
\end{equation} 
the ``nonextensive'' feature we are looking for.   The generalized 
homogeneous function works.

\begin{quote}
\begin{exercise}
 Why cannot we choose $F(T-T_c, L) = b^{-d+y_f} F(T-T_c, bL)$  as
       a simple generalization? Hint: non-extensive everywhere.
\end{exercise}
\begin{exercise}
 Is there any other generalization (other than
        Eq. \ref{eq:14} ) that could have been done?
\end{exercise}
\end{quote}

\subsubsection{Comments}
\label{sec:comments}

A few things are to be noted in these manipulations. 
\begin{multicols}{2}
\begin{itemize}
\item { A singularity of the free energy in the thermodynamic limit is manifest
in Eq. \ref{eq:16} and the singular behaviour is a power law in the
deviation from the critical point. These powers are called exponents.}
\item { We at the end
recovered the much coveted extensivity, everywhere save the critical
point, but at the cost of a scaling of the intensive variable.} 
\item What we are focusing here is the singular part of the
  free-energy.  There could ( and will) be an analytic piece that will 
  not show any such anomalous scaling.
\item{
With a positive $y_t$, we realize that the effective temperature is further
away from the critical temperature as the size is increased, if we
want to keep the free energy per particle ( or per unit volume) the
same. This desire to  keep the same  free energy per unit volume is
consistent with the content of Eq. (\ref{eq:4}). }
\item  A more dramatic result is that 
an argument ${L/ |t|^{-1/y_t}}$ can be thought of as a comparison of the 
length or size of the system with a characteristic scale $\xi$  of the system.
This scale $\xi$ diverges when $t \rightarrow 0$ as 
\fbox{$\xi \sim |t|^{-\nu}$} 
where \fbox{$\nu =1/ y_t$}.  
\item A diverging length scale is the hallmark of a
critical point and in fact {\it a critical point or criticality is
defined  as a point with a diverging length scale}.
\item No, we have neither devised a way of bypassing the Legendre
  transformation problem nor ignored it.  We chose the right ensemble
  (i.e. right thermodynamic potential) to do our analysis.  It is
  conceptually helpful to use the ensemble that involves the intensive
  variables.  For example, in first-order transitions (like, say,
  solid-liquid transition) there would be discontinuities in some
  extensive variables like the density, the energy density, the 
  entropy ( latent heat) etc, but the intensive variables remain
  continuous.  It is this fact that prompts one to use vertex
  functions in field theoretic analysis - but that's beyond the scope
  of these notes.
\item  It still begs the question:  What is this length $\xi$?
\end{itemize}
\end{multicols}

\subsubsection{Specific heat: power laws and exponents}
\label{sec:specific-heat}

The specific heat is $c \sim \partial^2f/\partial t^2 $.  From Eq.
(\ref{eq:16}), choosing $b=|t|^{-1}$, it follows that 
$c \approx f''(\pm 1)\ |t|^{-2+ y_f/y_t}$, where prime denotes
derivatives and $\pm$ stands for $t{>\atop <}0$.  Note that the
specific heat shows  a power law singularity with an exponent
\begin{equation}
\label{eq:17}
\alpha=2-  \frac{y_f}{y_t},\quad {\rm  defining}\ 
\alpha\  {\rm via}\ c \ \sim |t|^{-\alpha}, \ {\rm or}\
\alpha=-\ \lim_{t\rightarrow 0}\frac{\log c}{\log |t|}.
\end{equation}
The exponent for divergence of $c$ is same on both sides of the
critical point at $t=0$, though the amplitudes $f''(\pm 1)$ need not
be the same.

For $L<\infty$, the specific heat behaviour is 
\begin{equation}
  \label{eq:42}
\fbox{$  c(t,L)= L^{\alpha/\nu} \ {\cal C}(t L^{1/\nu}),$}
\end{equation}
making the $L$-dependence explicit as required by Eq. (\ref{eq:13}).
We leave it to the reader to establish the relationship between ${\cal
  C}$ of Eq. (\ref{eq:42}) and ${\cal F}$ of Eq. (~(\ref{eq:15}).

\begin{quote}
\begin{exercise}
The two-dimensional ferromagnetic Ising model, Eq. (\ref{eq:39}),  in
zero field ($h=0$) has a logarithmic divergence of specific heat, 
$c \sim - \ln |t|$.  How will the above formulation handle this?
Note also that $\ln x = \lim_{n\rightarrow 0} (x^n-1)/n$.
\end{exercise}
\end{quote}

\subsubsection{On exponents: hyper-scaling}
\label{sec:About-exponents}

Since free energy and energy-density cannot diverge, we need to have 
$\alpha < 1$. This puts a limit
\begin{equation}
\label{eq:18}
{y_f\over  y_t } \geq \ 1 \ {\rm or} \ \ y_f \geq  y_t  \ {\rm or} \ 
\nu \ \geq {1\over y_f}.
\end{equation}

At the critical point $k_BT_c$ is the energy scale, and therefore 
$F/(L^d  k_BT_c)$ dimensionally behaves like inverse volume. In
absence of any other scale, one might  expect
\begin{equation}
\label{eq:20}
\frac{F(t=0,L)}{ k_B T_c N} \ \sim \ L^{-d}, \quad \Longrightarrow
\fbox{$y_f=d$}. 
\end{equation}
A similar argument for $L \rightarrow \infty$, and $t \not = 0$,
 would then give, with $\xi$ as the important length-scale,
\begin{equation}
\label{eq:21}
{f\over k_BT_c} \ \sim \ \xi^{-d} \sim t^{+d\nu}, \quad
\Longrightarrow \fbox{$2-\alpha = d \nu$}. 
\end{equation}
This relation between $\alpha$ and $\nu$ involving $d$ is called 
{\it hyper-scaling}. Putting all together
\begin{equation}
  \label{eq:26}
  {f\over k_BT_c} = L^{-d}\, {\cal F}\left (\frac{t}{L^{-1/\nu}}\right ). 
\end{equation}
 In general
\begin{equation}
\label{eq:22}
2 - \alpha = y_f \, \nu
\end{equation}
where $y_f$ could be different from $d$ if some other length-scale plays
an important role at the critical point.

We did get the value of $y_f$ easily under certain conditions, but it
is not possible to obtain $\nu$.  Needless to say, with other
variables, there will be more exponents. The values of these exponents
are to be determined either from experiments or from statistical
mechanical calculations with appropriate Hamiltonians.

\begin{quote}
\begin{exercise}
If there is no singularity for  a finite system, then how can there be a
diverging length scale in say Eq. (\ref{eq:26})?  How to interpret
this properly? Hint: asymptotic expansion (see Ex V.5 and Eq. \ref{eq:59}).
\end{exercise}
\begin{exercise}
A problem from school algebra: Consider the sum $S(x,N) =
\sum_{n=0}^{n=N-1} x^n$ for $x\leq 1$. Show that for large $N$, and
$x$ close to $1$, $S(x,N)=N \tilde S (Nt)$ where $t=1-x$, and $\tilde
S(z)=(1-e^{-z})/z$. Recover, from this asymptotic form, the exact
result {\sf (i)} for $N\rightarrow\infty$ for any $x$ (Path b of
Fig. 1), and {\sf (ii)} for $x\rightarrow 1$ for any finite $N$ (Path a of
Fig. 1).  Compute numerically $S(x,N)$ for various values of $N$ and
$x\leq 1$, plot $S(x,N)/N$ vs $N t$ and compare with $\tilde S(z)$
(Path c of Fig. 1). Note the strong deviation from the asymptotic
scaling for small $N$ or large deviation of $x$ from $1$.  This is
actually a comparison of the partial sums of the infinite series
around the singular point.
\end{exercise}
\end{quote}

\subsubsection{Role of fluctuations: upper critical dimension: I}
\label{sec:Role-fluctuations}

From the form of the free-energy, the finite-size scaling of the energy
at $T=T_c$ can be obtained, namely $E/N \sim L^{-(1-\alpha)/\nu}$. Let
us now go back to Eq. (\ref{eq:10}).  The relative width of the
distribution for $E$ is given by
\begin{equation}
  \label{eq:57}
  \frac{\Delta E}{E} \sim L^{\alpha/2\nu} L^{(1-\alpha)/\nu}
  L^{-d/2}=L^{(y_f-d)/2}, 
\end{equation}
remembering that $N=L^d$, and using Eq. (\ref{eq:22}).  If now
hyper-scaling is valid, $y_f=d$, and see, the relative width does not
vanish even in the limit of $L\rightarrow\infty$.  Therefore for such
a case, fluctuation plays a crucial role.  If, however, $y_f < d$,
then the relative width vanishes in the thermodynamic limit.  We then
get a situation where we have a critical point with a diverging
length-scale and possibly diverging response functions, but the effect
of fluctuations is not strong in the sense that the probability
distribution remains sharp around the average value.
Similar results are found for other extensive variables also as shown
in Sec \ref{sec:Role-fluct-upper}.  

One thing becomes clear: the dimensionality of the system is very
important and the lower the dimensionality the stronger is the effect
of fluctuations.  We might expect that for large enough $d$, this
condition $y_f<d$ will be satisfied.  One may then try to understand
such critical systems by considering the average system ignoring
fluctuations altogether.  This class of theories is called mean-field
theory. 

If there is a finite value of $d=d_u$ above which $y_f < d$, then
fluctuations can be ignored for all $d>d_u$.  This $d_u$ is called the
{\it upper critical dimension} of the system.  Hope would be that the
effect of fluctuations can be studied in a controlled manner around
this $d_u$ with $\epsilon=d_u -d$ as a small parameter.  This is the
basis of the $\epsilon$-expansion for many systems.

We, however, caution the reader that the finite-size scaling in a 
mean-field theory ( or $d>d_u$)  is more complicated.
We are concentrating  only on the fluctuation dominated case.

\subsection{Solidarity with  thermodynamics}
\label{sec:faith-thermodynamcis}

The problem with extensivity can now be understood in terms of a
diverging length scale.  In the strict sense, there is actually no
violation of extensivity.  The idea of adding small pieces to build a
bigger one makes sense provided the smaller pieces themselves are
representative of the bulk.  For systems with short range interactions
and small intrinsic or characteristic length scales, this is
reasonable because the ``small size'' effects or boundary effects are
perceptible only if the size is comparable to these lengths.  These
effects are small corrections that can be safely ignored.  In case of
a diverging length-scale or characteristic length-scales larger than
the system size, smaller pieces cannot be added up.  Thus at a
critical point with diverging length-scales, the notion of adding up
smaller pieces fails.  The length-scale $\xi$ is the appropriate scale
for comparison and so at a critical point the whole sample is to be
treated as a single one.

At any temperature $t\not = 0$, $\xi \sim |t|^{-\nu}$ is large but
finite.  If we take a block of size $\xi^{d}$ as a unit, then
extensivity  requires that the free energy be proportional to the
number of such blocks or blobs  i.e.  
$F = (N/\xi^d)\, f_0 \sim N |t|^{d\nu} f_0$, with $f_0$ as the
free-energy  per blob.  If $f_0$ is independent of $\xi$, then 
$f \sim f_0\, |t|^{d\nu}$, recovering Eq.~(\ref{eq:21}).  In case
$f_0$, the free energy of a blob of volume $\xi^d$, depends on $\xi$,
then extra $\xi$ contribution is expected and one gets
Eq. (\ref{eq:22}) with $y_f\not = d$. 

\begin{quote}
\begin{exercise}
Anisotropic system: A particular two-dimensional system of size
$L_x\times L_y$ shows the following scaling behaviour
\begin{equation}
  \label{eq:56}
  f(t,L_x,L_y)\approx X {\sf f} \left( L_xt^{\nu_x}, L_yt^{\nu_y}\right),
\end{equation}
$X$ being the size dependent prefactor.  There are now two different
length scales in the two directions.  What are the length-scale
exponents? Find the relation (hyper-scaling) among $\alpha,\nu_x$ and
$\nu_y$.  What would be the form of $X$ if one takes $L_x\times\infty$
strips. (Ans: $X=L_x^{\alpha/\nu_x}$.)\ \ What would be $X$ for
$\infty\times L_y$ strips?  We assume that the strips do not show any
phase transition.  What would be the right combination variable if
both $L_x$ and $L_y$ are to be used in the prefactor?  (Ans: 
$X=(L_x^{-1/\nu_x}+ L_y^{-1/\nu_y})^{-\alpha}$. Why?)
\end{exercise}
\begin{exercise}
Cross-over:  Think of a cylindrical geometry.  The system is
finite in one direction (length $L$)  but infinite in the remaining $d-1$
directions.  There will be a  critical behaviour in this geometry
if $d$ is not too  small.  Consider the $d-1$ to $d$ dimensional
crossover as the finite length $L$ is made larger.  Consider various
situations: (a) $d < d_u$, (b) $d>d_u$ but $d-1<d_u$, and (c)
$d-1>d_u$.
\end{exercise}
\end{quote}

\subsection{More Variables: temperature and field}

\label{sec:more-variables}

Generalization of the scaling of Eq. (\ref{eq:14}) to more variables
is straight forward.  Take a variable $h$ which represents another
intensive quantity like pressure, magnetic field etc, measured from
its critical value. The critical point is now at $t=0,h=0$.  The free
energy can be written as
\begin{equation}
  \label{eq:25}
  F(t,h,L) = b^{-d+y_f} \ F(b^{-y_t} t,b^{-y_h} h, bL),\quad
  {\buildrel {N\rightarrow\infty}\over \Longrightarrow} f(t,h) =
  b^{y_f} f(b^{-y_t}   t,b^{-y_h} h),
\end{equation}
so that the same series of manipulations done for $h=0$ would lead to 
\begin{mathletters}
\begin{eqnarray}
\label{eq:23}
{\sf f}(T,h,L)\equiv \frac{1}{N} F (t,h,L) &=& \xi^{-y_f} {\cal
  F}_{\pm}\left ({h\over\xi^{-y_h}},{L\over \xi}\right),\\
\label{eq:23b}
&=& L^{-y_f} \tilde{\cal F} \left(
  \frac{t}{L^{-1/\nu}},\frac{h}{L^{-y_h}}\right), 
\end{eqnarray}
\end{mathletters}
where ${\cal F}_{\pm}(x,y) = F(\pm 1,x,y)$.

If $tL^{y_t}=const $, then, as per Eqs. (\ref{eq:23b}) and (\ref{eq:26}), 
extensivity is nowhere to be found - 
  that's curve (c) in Fig 1.  The scaling form of Eq. (\ref{eq:26}) or
(\ref{eq:23b}) is known as finite-size scaling. 

\npg
\subsubsection{Comments}
\label{sec:Comments}

\begin{multicols}{2}
\begin{itemize}
\item For a fixed $N$, if $t\rightarrow 0,$ then ${\sf f} \sim L^{-y_f}$. 
\item For a
fixed $t$, if $N\rightarrow\infty,$ then 
\begin{equation}
  \label{eq:38}
f \sim |t|^{2-\alpha}\,
\hat{\cal F}_{\pm}\left(\frac{h}{|t|^{y_h/y_t}}\right). 
\end{equation}
\item We see the general feature: \\
critical region $\Longleftrightarrow $ \ ``nonextensivity'' \
$\Longleftrightarrow $ \  
finite size scaling \ $\Longleftrightarrow $ \ Scaling.
\item Once we take the diverging length-scale as the sole scale for the
problem, the finite-size scaling form can be obtained from the bulk
behaviour as well.  A finite length $L$ matters only when it is
comparable to $\xi$.  For $L\gg\xi$, the above-mentioned blob picture
is valid and for $L<\xi$ the whole system needs to be treated as a
critical one.  If the bulk behaviour is like 
$c(t) \sim |t|^{-\alpha}$, then the size-dependent cross-over is given
by 
${\sf c}(t,L) \sim |t|^{-\alpha} \ \tilde{\cal C}(L/\xi, h/\xi^{y_h})$
$ \sim L^{\alpha/\nu} \ {\cal C}(t/L^{-1/\nu}, h/L^{y_h})$. (See
Eq. (\ref{eq:42}).)   Historically, bulk scaling was
proposed  first and  finite-size scaling came later on.  
\item We cannot overemphasize the fact that $h$ and $t$ are completely
  independent variables. Yet in Eq. \ref{eq:38} the free-energy in
  the thermodynamic limit depends on the single combination variable
  $h/t^{y_f/y_t}$, and not on $h$ and $t$ separately.  Results of
  experiments (real or numerical) done at various values of $h$ and
  $t$ can be collapsed on to a single curve, unthinkable away from the 
  critical region.
\item It seems there is one exponent too many for the bulk free energy 
  density $f$ in Eq. (\ref{eq:25}), $y_f/y_t$ and $y_h/y_t$ would have 
  sufficed.  Yes, it would if we consider thermodynamics alone.  But
  the path via finite-sized system showed us the necessity of another
  exponent, $\nu$, and so we keep all the three in Eq. (\ref{eq:25}).
\end{itemize}
\end{multicols}

\begin{quote}
\begin{exercise}
Take the Legendre transform of $f$ in Eq. (\ref{eq:25}) with respect
to $t$.  Discuss its scaling property.  Caution: The scaling is not for the
total entropy or entropy density. The special scaling is for the
deviation from the critical value of the entropy at the critical
point.  This shows the difference of the scaling around the critical
point and the simple scaling of thermodynamics.
\end{exercise}
\end{quote}

\subsubsection{More variables mean more exponents}
\label{more-variable}

The power laws we saw earlier tend to suggest similar behaviour for
other physical quantities also. We define several such practically
important exponents, but shall ultimately see that all of these can be 
expressed in terms of the three $y_f,y_t, y_h$
introduced earlier.  

For concreteness, the magnetic language of the Ising model of Eq.
(\ref{eq:39}) is used.  Let us assume that there is a critical point
for the Ising model (and there is one) so that the free energy is
given by Eq, (~\ref{eq:38}).  We have already defined $\alpha$ as the
specific heat exponent. We define $\beta$ for magnetization, $\gamma$
for susceptibility, $\delta$ for critical isotherm.  Apart from these
thermodynamic exponents, two other expoents are needed $\nu$ ( already
introduced) for length-scale and $\eta$ for critical correlations.

Remembering that the magnetization is the derivative of the
free-energy with respect to $h$, one gets 
$m\approx |t|^{(y_f-y_h)/y_t} {\cal F}_{\pm}\!'(h/t^{y_h/y_f})$.  From
this, see that for $h=0$, the magnetization vanishes at the critical
point as 
\begin{equation}
  \label{eq:8}
  m\sim |t|^{\beta}, \quad {\rm with}\quad \beta=(y_f-y_h)/y_t.
\end{equation}
For a ferromagnetic transition, there is no magnetization in zero
magnetic field in the high temperature phase (paramagnet), and
therefore ${\cal F}_{+}\!'(0)=0$ but ${\cal F}_{-}\!'(0)\not =0$.
A quantity like $m$ that describes the ``ordering'' of the system is
called an order parameter and $\beta$ is the order-parameter exponent.
We repeat that phase transitions need not necessarily have an order parameter
(see Sec. \ref{sec:Example:-Polymers}).

Susceptibility $\chi$ is the response of
magnetization, and  we see in zero field ( $h=0$) ($\chi = \partial
m/\partial h$)
\begin{equation}
  \label{eq:40}
  \chi \sim |t|^{-\gamma}, \quad {\rm with} \quad \gamma=(2y_h-y_f)/y_t.
\end{equation}

At  the critical point ($t=0$) in presence of a field ( critical
isochore ), we  get 
\begin{equation}
  \label{eq:27}
  m\approx h^{1/\delta} m_0, \quad \delta= {y_h/(y_f-y_h)}.
\end{equation}
Here the amplitude 
$m_0= \lim_{x\rightarrow \infty} x^{-(y_f-y_h)/y_h} {\cal  F}'_{\pm}(x)$. 
In general, $\delta \not = 1$.  One expects a
linear relationship (``linear response'') between ``cause'' ($h$) and
``effect'' ($m$), but that turns out not to be the case at
criticality.  A linear relation can never give an infinite $\chi$!

A major consequence of the homogeneity of the free energy is the power 
law behaviour of various physical quantities (various derivative of
free energy) and all of these are
obtained from the three basic exponents.   In case of hyper-scaling
(i.e. $y_f=d$) we have a further reduction and only two exponents are
needed for a complete description of the critical behaviour of a
system.

For a thermodynamic description this looks enough, but we already saw
the usefulness of a length-scale based analysis.  We need to define
the length-scale properly and in the process we will find a more
useful critical exponent $\eta$.

\subsubsection{Comments}

\begin{multicols}{2}
\begin{itemize}
\item The blob picture can be used for susceptibility near the
  critical point. There are $N/\xi^d$ blobs. Each blob of size $\xi^d$
  can be thought of as a critical object.  The total susceptibility is
  given by $N\chi=(N/\xi^d) \chi_{\rm blob}$, where $\chi_{\rm blob}$
  is the total susceptibility of a blob.  Finite-size scaling predicts
  $\chi_{\rm blob}=\xi^d \xi^{\gamma/\nu}$ so that we obtain $\chi \sim
  |t|^{-\gamma}$
\item Note that we require $y_h <y_f<2y_h$.
\end{itemize}
\end{multicols}

\subsubsection{Role of fluctuations: upper critical dimension:II}
\label{sec:Role-fluct-upper}

The role of fluctuations was analyzed earlier in the context of
specific heat.  Let us now reanalyze it for the probability
distribution for magnetization $M$.  The finite-size scaling behaviours
of (total) magnetization and (total) susceptibility at $T=T_c$ are 
$\langle M\rangle\sim L^d L^{-\beta/\nu}$ and 
$N \chi\sim L^d L^{\gamma/\nu}$.
The relative width $\Delta M/M$ of the probability distribution for $M$ is
$L^{\gamma/2\nu} L^{\beta/\nu} L^{-d/2} = L^{(y_f-d)/2}$ 
where Eqs. (\ref{eq:49}) and (\ref{eq:22}) have been used.

It is reassuring that the condition for sharpness (or broadness) of
the probability distribution for $M$ is the same as obtained for
specific heat (see Sec. \ref{sec:Role-fluctuations}).  The upper
critical dimension is the same no matter which conjugate pair we use.

\subsection{On exponent relations}
\label{sec:More-expon-relat}

Once the exponent identifications are made, with only two independent
ones,  it is possible to write down many relations involving the
(in principle) experimentally measurable exponents. For example, by
adding the exponents,
\begin{equation}
  \label{eq:49}
  \alpha + 2\beta+\gamma=2.
\end{equation}

If we take the free-energy density  $\sim h m$ and $m = \chi^{-1} h$, then it
follows that $2\beta+\gamma = 2- \alpha$, as in the above equation,
without using any explicit formula.  Another way of re-writing the
above relation $(-\alpha=2(\beta-1) -\gamma)$ suggests that the
behaviour of specific heat ($c_{h=0}$)  
is similar to $(\partial m/\partial t)^2 \chi^{-1}$.  In fact,
thermodynamics gives us the formula 
\begin{equation}
  \label{eq:50}
 c_h - c_m = T \left(\frac{\partial m}{\partial T}\right)_h^2 \chi_T^{-1},
\end{equation}
where $c_x$ is the specific heat with $x$ constant.  Since specific heat is
positive definite, it follows from Eq. (\ref{eq:50}) for $c_h$ 
with the intensive variable $h$ held constant at the critical value $h=0$ that 
$\alpha + 2\beta + \gamma \geq 2$.

\begin{quote}
\begin{exercise}
Thermodynamic argument seems to indicate inequality rather than
equality in Eq. (\ref{eq:49}).  Find out the conditions for which
a strict inequality is expected.
\end{exercise}
\end{quote}

\npg 
\section{Relevance,  irrelevance and Universality}
\label{sec:Relev-irrel}

A few observations should not miss our attention.  Since $y_t>0$, the
combination variable $Lt^{1/y_t}$ in, say, Eq. (\ref{eq:42}) or
(\ref{eq:23b}) have different limits for $t=0$ and $t\not= 0$ in
the limit $L\rightarrow \infty$.  This difference actually gave us the 
different $L$-dependent behaviour of the free-energy or the specific
heat, or, as a matter of fact, any physical quantity we may calculate or
observe.   In the same way, if $y_h>0$, then in the bulk case, if
$t\rightarrow 0$, i.e. as the critical point is approached, the
combination or scaling variable $h/t^{y_h/y_t}$ goes on increasing if
$h\not= 0$, no matter how small it is.  The behaviour is 
different if $h$ is strictly zero.  Such a tendency of a parameter to 
grow also tells us that zero field and nonzero field behaviours are
different.  We call these variables {\it relevant variables}.

In case $y_h <0$, then for $t\rightarrow 0$, the scaled variable is
zero irrespective of its value.  In such a case whether the system has
$h\not= 0$ to start with  is immaterial.  No wonder these are to be called  
{\it  irrelevant variables}.

There could also be {\it redundant} terms that don't matter at all,
like a constant added to a Hamiltonian. We ignore them altogether.

To be at a critical point, the relevant variables must be tuned
properly to be at their critical values (e.g. $t=h=0$ in the magnetic
example) because they take us away from the critical point.
Irrelevant variables don't matter as such  but they do play  
a significant role if we want to go beyond the leading behaviour.

Special situations arise, if the scaled function shows a singularity
as an irrelevant variable scales to zero.  Such variables are then
important for the critical behaviour, though they don't take the
system away from criticality.  Such a variable is called a {\it dangerous}
irrelevant variable.

Critical points are classified by the number of relevant variables
required to describe them.  An ordinary critical point requires two (for 
the magnetic problem temperature and magnetic field, for the
liquid-gas transition temperature and pressure).  If three relevant
parameters are needed it is called a tricritical point and so on.

 Our analysis so far has been restricted to thermodynamic parameters
like $t, h$ etc, but this can be extended to any parameter occurring
in the Hamiltonian in a statistical mechanical approach.  The starting
Hamiltonian may have a large number of parameters based on the
microscopic details of the system.  But as we look at longer
length-scales close to the critical point, all these parameters can be
classified under the banner of relevance and irrelevance.  By throwing
away the irrelevant terms, for the leading behaviour, an enormous
simplification ensues (e.g, all ordinary critical points will have
only two relevant parameters, only difference may be in the numerical
values of the exponents $\nu$ and $\eta$, etc.).  It might then be
expected that the numerical values of the exponents could be
identified from certain basic symmetries etc of the Hamiltonian. 
This is the concept of Universality.  A universality class would be
described  by the exponents and also the amplitude ratios of the
various singular quantities on the both sides of the critical point.

The idea of universality transcends the domain of critical phenomena.
Whenever we are interested in properties on a scale much bigger than the 
underlying or microscopic scales, there seems to be a set of properties which
are quantitatively same no matter what the microscopic details are.  The  
exponents we have seen are just one such examples.  Historically it was found 
that the shape of the coexistence curve near the liquid-vapour critical point
is independent of the chemical composition and is also very similar to the
critical behaviour of several magnets.
Once the universality class of a system is identified, the set of universal 
quantities can be obtained by studying a simpler model system than the original
one with all details. The simpler model is expected to focus on the relevant 
variables only or at most a few irrelevant ones.

Universality is not just a set of exponents.  In a scaling description, 
the amplitudes and even the scaling functions are independent of gross 
details except that the arguments of the functions may involve nonuniversal
metric factors. In a finite size scaling there could be a dependence on the 
boundary conditions as well.   

Whether a variable is relevant or not is determined by its scaling exponent 
(e.g. $y_h$ in the previous case).
In certain cases, one may get these by simple arguments (Gaussian model in 
Sec. \ref{sec:models-as-examples}) but in most cases these are to be 
determined.
\begin{quote}
\begin{exercise}
Can there be critical cases with only one relevant variable or no relevant
variables at all? 
\end{exercise}
\end{quote}

\npg
\section{Digression: first-order transition and transition with
no ordering}  
\label{digression}

\subsection{A first-order transition: $\alpha$=1}
\label{sec:first-order-trans}

What happens at the borderline of $\alpha=1$?  The form of the
free-energy tells us that the energy-density will have a discontinuity
on the two sides of the singular point.  Such a phase transition with
a discontinuity in any first derivative of the free-energy is defined
as a first-order transition.  This value of $\alpha$ leads to
$\nu=1/d$.

We see the possibility of a first order transition in the same
framework developed for the critical point.  But first-order
transitions can be of other types also, and they may not
necessarily have any  diverging length-scales associated with
it.  One needs to be careful about it.

As an example, let us consider a very simple  configuration space for a 
magnet (a crude approximation to an Ising-type model).  We replace the
$N$-spin configurations by two types only.  
A zero-energy ground state --  all spins up or all down -- two-fold
degenerate, and an excited state of energy $N\epsilon$ with
degeneracy $2^N$.  (This is a one-dimensional ferro-electric 
six vertex model, in disguise.)  The partition function for this model 
is $Z(x,N)= 2 + (2x)^N$ where $x=\exp(-\beta\epsilon)$.  It is easy to 
see, by taking the $N\rightarrow\infty$ limit, that there is
first-order transition at $x=1/2$ with total energy going from zero
for $x<1/2$ to $N\epsilon$ for $x>1/2$. No problem with extensivity
for $x\neq 1/2$.  In the limit $N\rightarrow\infty$, the specific heat 
is just a delta function at the transition point. For finite $N$,
the specific heat  can be written in the form 
\begin{equation}
  \label{eq:59}
 C(x,N)=N^2\  \ln 2 \ \frac{2e^{Nt}}{(2+e^{Nt})^2},   
\end{equation}
where $t=1-2x$.  For a $d$-dimensional system, take $N=L^d$. What we
now see from the scaling variable that there is a diverging length
scale with exponent $\nu=1/d$.

\subsubsection{Comments}
\label{sec:Comments-3}

\begin{itemize}
\begin{multicols}{2}
\item Fig 2 shows $C(x,N)$ for various values of $x$ and $N$ computed
  from the partition function and compared with the scaling function
  of Eq. \ref{eq:59}. This is an example of {\it data collapse}.
\item  The thermodynamic limit
  for any given value of $x$ (Path b of Fig. 1) comes from the tail
  of the scaling function of Fig 2.  This plot amplifies the critical
  region of Fig. 1. 
\item The $N^2$ prefactor in Eq. \ref{eq:59} is consistent with an
  exponent $\alpha/\nu$ with $\alpha=1$.  Remember that $N=L^d$.
\item The peak of the scaling function in Fig. 2 is not at the bulk
  transition point.

\begin{quote}
\begin{exercise}
What is this length scale in the context of this model?
\end{exercise}
\end{quote}

\vbox{
\begin{figure}[htbp]
  \begin{center}
    \psfig{file=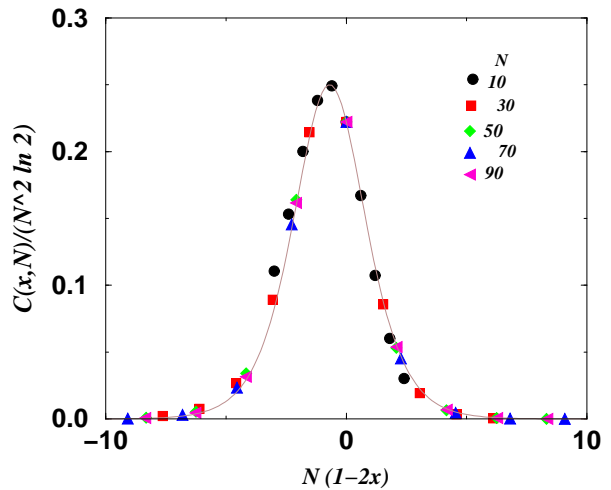,width=3.1in}
\narrowtext
    \caption{Data collapse of $C(x,N)/N^2$ vs $N t$ for various $N$.
      The solid line is the finite-size scaling function of
      Eq. \ref{eq:59}.}
    \label{fig:2}
  \end{center}
\end{figure}
}
\end{multicols}
 
\end{itemize}
\widetext

This simple model calculation can be generalized.  One way of
studying a first-order transition is to compute the free energies of the two 
phases independently ( like e.g. solid and liquid) and then finding the lower
free energy curve.  Taking these free energies per particle, $f_1$ and $f_2$,
as coming from a restricted sum over states of the full partition function,
the total partition function can be approximated by 
$Z=\exp(-N\beta f_1) + \exp(-N\beta f_2)$, with the free energy per particle
$f=\min(f_1,f_2)$ for $N\rightarrow\infty$.  In general, $f_i$'s are analytic 
function and the transition temperature is determined by $f_1(T_c)=f_2(T_c)$,
so that $\Delta f = |f_1(T) -f_2(T)| \sim |T-T_c|$ from a Taylor series 
expansion around $T=T_c$.  The partition function can therefore be written as
$Z=\exp(-N\beta f) \ [ 1 + \exp(-N/\xi^d)]$, where 
$\xi^{-d} = \beta_c|\Delta f|$.
This defines a length scale $\xi \sim  |T-T_c|^{-1/d}$. 
For first-order transitions, {\fbox{$\nu=1/d$}}.

\subsection{Example: Polymers : no ``ordering"}
\label{sec:Example:-Polymers}

We give an example here of a phase transition for which there is no
``ordering'' unlike the magnetic case, and the interest in the
critical point is because  of finite-size behaviour.

Polymers are long linear objects, abundantly occurring in nature.  Let
us take a single flexible polymer in a solvent ( a single polymer of
length $N$ in $d$-dimensional space of infinite volume).  Depending on
the nature of the solvent and the monomers constituting the polymer,
there could be effective repulsion or attraction between the monomers.
By changing temperature, one may go from a repulsion-dominated
(monomer favouring solvent molecules) (self avoiding walk) to an
attraction-dominated (monomer preferring monomer) phase.  This is
called a collapse transition.  Such a transition for a single molecule
can occur only in the thermodynamic limit of length
$N\rightarrow\infty$.  But that's of no interest because the polymers
are necessarily of finite lengths.  The two phases in this example are
described by the overall size of the polymer, e.g. by the mean square
end-to-end distance $\langle R^2\rangle\sim N^{2\nu}$. In a collapse
transition the value of this size exponent $\nu$ changes from
$\nu\approx 3/(d+2)$ to $\nu=1/d$ in $d$-dimensions.  This is an
example of a phase transition or critical point which may not be
viewed as ordering (unlike the magnetic system) - no less important
though.  The main features of the transition are reflected in the
``finite-size'' behaviour (with respect to $N$) both at the transition
and in the two phases, and the general analysis we have done so far is
easily applicable here.

\begin{quote}
\begin{exercise}
What sort of a critical point is the collapse transition? 
Hint: How many relevant parameters  at a collapse transition?
We need to make length going to infinity, we need to adjust temperature 
and if we take a solution of polymers, we need to make concentration
equal to zero ( i.e. single chain limit).\\  Ans: tricritical point.
\end{exercise}
\begin{exercise}
For a polymeric phase transition of the type  mentioned above, show that 
the specific heat   
per unit length (or per monomer) can be written in a scaling form
$c= N^{2\phi -1} {\cal C}( (T-T_c) N^\phi)$ 
\end{exercise}
\end{quote}
\npg

\section{Exponents and correlations}
\label{sec:Exponents-relations}

\subsection{Correlation function}
\label{sec:Correlation-function}

Consider the Ising model of Eq. (\ref{eq:39}) or ~(\ref{eq:37}) and
take $M=\sum_i \langle s_i \rangle$.  The fluctuation-response theorem 
of Eq. (\ref{eq:11}) and translational invariance can be used to
express the susceptibility ($\chi$ per particle)  as
\begin{equation}
  \label{eq:41}
  \fbox{$N\chi = (k_BT)^{-1} \sum_i\sum_j \langle (s_i - \langle
s\rangle) (s_j - \langle s\rangle) \rangle = N(k_BT)^{-1} \sum_{\vec x}
g({\vec x})$},
\end{equation}
where 
$g(\vec x)$ =
$\langle(s_0-\langle s\rangle)(s_{\vec x}-\langle s\rangle) \rangle$ 
is the pair correlation function. The behaviour of the
susceptibility is therefore dependent on the behaviour of the
correlation function.   This relation is quite general as pointed out
in section \ref{sec:Comments-1}. By expressing the total quantity
$\Phi=\sum_{\bf x} \phi({\bf x})$, in terms of a local density, we have 
$\langle\Phi^2\rangle - \langle\Phi\rangle^2$ 
$ =N \sum_{\bf x} g_{\phi\phi}({\bf x})$, where 
$g_{\phi\phi}({\bf x})=$ 
$ \langle (\phi(0)-\langle\phi\rangle)(\phi({\bf x}) -\langle\phi\rangle)
\rangle$.

For simplicity, let's replace the sum in Eq. (\ref{eq:41}) by an
integral so that $\chi = \int d^d x g(\vec x)$.  For the Ising
case, $s$ is bounded and so  is $g$.
We also know that $\chi$ diverges at least at $T_c$.  Only way this
can happen is from the 
divergence of the integral or the sum - it is the large
distance property of $g(x)$ that controls the behaviour.  We can
conclude that, at least at $T_c$, $g(x)$ cannot be a short
ranged function, but it has to decay to zero  for infinitely large
distances.  In $d$ dimensions,
 the decay at $T=T_c$ for large $x$ is given by
\begin{equation}
  \label{eq:45}
\fbox{$g(\vec x) \sim x^{-(d-2+\eta)}$}\quad {\rm  with}\quad  \eta
\geq 0\quad \quad (T=T_c).  
\end{equation}
For $T\neq T_c$, convergence requires that $g(r)$ should decay
sufficiently faster than this, better be of short-range character with
a characteristic length scale $\xi$, e.g.  $\exp(-x/\xi)$.  The decay
for any temperature can be written as
\begin{equation}
  \label{eq:46}
g(\vec x) = \frac{1}{x^{d-2+\eta}}\ {\sf
  g}\left(\frac{x}{\xi}\right)\quad \quad (T\not= T_c), \quad\quad
{\rm(see\ comments\  below)}
\end{equation}
which goes over to the $T=T_c$ case if this length-scale $\xi$
diverges for $T\rightarrow T_c$.  The length-scale we saw popping out
naturally for comparisons of the length of the system can now be
identified as the correlation length, the scale for correlations.  A
precise definition\footnote{ There are in fact many ways of defining a 
length scale. We use a $g(x)$-based definition.} 
of $\xi$ would be from the second moment of
$g({\vec x})$ as
\begin{equation}
  \label{eq:43}
  \xi^2 = \frac{\int d^d { x}\ x^2 g({\vec x})}{\int d^d { x}\
    g({\vec x})} \sim |t|^{-2\nu} .
\end{equation}
(For an isotropic system, the first moment vanishes by symmetry.)  It
is possible to define such scales via higher moments also.  Under the
assumption of one scale, all of these will have similar divergences at
a critical point.  Nevertheless, it is worth keeping in mind that
there are cases where one may need to study higher order moments and
different moments defining different length-scales.

\begin{quote}

\begin{exercise}
Can $\eta <0$?
\end{exercise}
\begin{exercise}
Generalize Eq. (\ref{eq:43}) for anisotropic cases like in Ex. V.7.
\end{exercise}
\end{quote}

\subsubsection{Comments}
\begin{multicols}{2}
\begin{itemize}
\item The on-site term, $i=j$, with prefactor $(k_{B}T)^{-1}$ in Eq.
  \ref{eq:41} is the susceptibility of individual spins (or isolated
  degree of freedom), the Curie susceptibility.  The correlation
  contribution is special to an interacting system.  Note that the
  correlation vanishes for a noninteracting system.
\item The fluctuation becomes long-ranged at the critical point.  On
  the high temperature side $\langle s_i\rangle=0$ ( no
  magnetization), and so the fluctuation correlation is the same as
  the spin-spin correlation.
\item For low temperatures ($T<T_c$), the spin-spin correlation
  approaches a constant ($\langle s\rangle^2\neq 0$) for large
  separations.  This approach to the constant value is generally
  short-ranged, i.e., very rapid like exponential but becomes
  long-ranged at $T_c$.
\item We get a long-range correlation even though the interactions are
just short ranged ( e.g., nearest neighbour interaction in the chosen
Ising model).
\item A finite size scaling for the correlation length itself would be  
 $\xi \sim L$.  Quite often it is possible to write it as an equality
with the amplitude determined by the known exponents. The amplitude  surely 
depends on which correlation function is used to determine $\xi$, boundary 
conditions etc.
\item For historical reasons Eq. (\ref{eq:46}) is generally written in a 
slightly different form as $g(\vec x) = \frac{1}{x^{d-2+\eta}}\ {\sf
  D}\left(\frac{x}{\xi}\right) \exp(-x/\xi)$, with $D(z) \rightarrow 1$ for
$z\rightarrow 0$, while $D(z) \sim z^{(d-3+2\eta)/2}$ for large $z$.  
Such a form is required
to make correspondence with the Ornstein-Zernicke theory of correlations, 
which we shall not discuss in these notes.
\end{itemize}
\end{multicols}

\subsection{Relations among the exponents}
\label{sec:Relations-among-the}

Let's think of the pair correlation function.
It decays rapidly once we are on a scale greater than the correlation
length $\xi$.  Close to $T_c$, we can think of the system as
blobs of highly correlated regions - the blobs are of size
$\xi^d$ in $d$ dimensions.   These are the blobs we introduced to
salvage extensivity in Sec. \ref{sec:faith-thermodynamcis}.
Inside a blob $(r<<\xi)$, the  fluctuations are critical-like
and at a simple level, a blob can be thought of as at $T_c$.
On a bigger length scale $x>>\xi$, the blobs are independent.
Basically, we are arguing that it is the correlation length that
matters - all other length scales are unimportant. 

We use this simple picture for the susceptibility.  We can cutoff the
integral at $x\sim \xi$, and inside this region 
$g(x) \sim x^{-(d-2+\eta)}$.  The integral 
$\int^{\xi} dx\ x^{1-\eta} \sim \xi^{2-\eta}$.  Using the temperature 
dependence of $\xi$, we get the temperature dependence of $\chi$ as 
$\mid t\mid ^{-\nu (2-\eta)}$.
The net result is
\begin{equation}
  \label{eq:44}
 \gamma = \nu (2-\eta). 
\end{equation}
This relation with the help of Eq. (\ref{eq:40}) gives 
\begin{equation}
  \label{eq:47}
  y_h=\frac{y_f+2 -\eta}{2}.
\end{equation}
This is a very important relation - it shows  how the external
magnetic field or the nonthermal relevant variable scales away from
the critical point is determined by the decay of correlations {\it at
criticality}.

It is straightforward to see that $\alpha=\nu(2-\eta_E)$.

\subsection{And it's correlations}
\label{sec:Its-correlations}

We  now have the full identification of the three exponents:
\begin{equation}
  \label{eq:48}
  \fbox{$y_t=1/\nu, \ y_f= d, \  y_h=\frac{d+2 -\eta}{2}$}. 
\end{equation}
All the nuances of critical phenomena are qualitatively and
quantitatively expressed in terms of the correlation function and the
exponents needed for it.  It's a shift from a thermodynamic
description to a purely statistical mechanical one.  {\it It is the
correlations of the degrees of freedom that control completely the
whole phenomenon.}

In this particular example of ordinary critical point, one of the
exponents, $\eta$ is really an exponent defined at criticality, while
the other,$\nu$, is an off-critical one.  An off-critical exponent helps in 
the description of the approach to the criticality.  This can however be 
treaded  for purely critical exponents. 

\begin{quote}
\begin{exercise}
What if $y_f\not = d$?
\end{exercise}
\end{quote}

\subsubsection{Comments}
\label{sec:Comments-4}
\begin{multicols}{2}
\begin{itemize}
\item 
Now that we have emphasized the role of correlation functions and
expressed $y_h$ in terms of $\eta$, why are we not doing the same
thing for specific heat?  Eq. (\ref{eq:10}) tells us that specific
heat is related to energy-energy correlation 
function, and so shouldn't we define an exponent $\eta_{\rm E}$?  We could
if we want to.  But convince yourself, by using hyper-scaling, that
 $y_t=1/\nu = (d+2 -\eta_{\rm E})/2$.   We may use either $\nu$ or
 $\eta_{\rm E}$.
\item  One may generalize the analysis of Sec.\ref{sec:Correlation-function}
for any local variable $L(\{\phi(x)\})$ and the corresponding response 
function will be determined by the critical correlation of  
$\langle L(\{\phi(x_1)\}) L(\{\phi(x_2)\})\rangle$.
\end{itemize}
\end{multicols}

In a correlation based approach, we need to compute the critical correlations
of all possible combinations of the degrees of freedom (e.g. in the Ising 
case, spin-spin, energy-energy and so on).  From these,  the relevant 
variables can be identified. In the Ising criticality case, there are only two.
In addition, for a thermodynamic problem,  we need to know the finite size 
behaviour of the free energy at the critical point (or assume hyperscaling).  
These three critical parameters then completely specify the leading behaviour 
even around the transition point.  
 
To repeat, a thermodynamic description would focus on $y_f,y_t,y_h$ for a 
critical point, while statistical mechanical description would focus on 
the set of purely critical exponents $y_f/y_t, \eta_{\rm E}$, and $\eta$.  
The scaling relations we derived show their equivalence.  

\begin{quote}
\begin{exercise}
Can this happen: the spin-spin correlation function is given by
Eq. (\ref{eq:46}) and the energy-energy correlation is given by 
$g_{EE} \sim x^{-(d-2+\eta_E)} {\sf g}_E (x/\xi_E)$ where 
$\xi_E \sim |t|^{-\nu_E}$, but $\nu_E\not= \nu$?
\end{exercise}
\end{quote}
 
\subsection{What's $b$ anyway? : length-scale dependent parameters}
\label{sec:Whats-b-anyway}

If we look at the free energy of Eq. (\ref{eq:16}) or
Eq. (\ref{eq:25}) in the thermodynamic limit of $L\rightarrow\infty$,
a question may be asked, ``what is $b$ now?''.  Answer is 
in the  correlation function.  Under a rescaling
$x\rightarrow x/b$, we see $g(x,t) =  b^{-(d-2+\eta)} g(x/b,b^{y_t} t)$, 
where the $t$-dependence has been made explicit.  This resembles the
scaling of the free-energy. 
An interpretation of this equation is that if we change the scale of
length measurements (actually the scale of microscopic details -
though not apparent right now), the parameters of the problem and the
concerned physical quantity  get rescaled.   Quantitatively, 
for $x\rightarrow x'=x/b$, we have  $t\rightarrow t'=b^{y_t} t$,
$h\rightarrow h'=b^{y_h} h$, and $f\rightarrow f'=b^{y_f} f$.
  
For a given problem (i.e. $t,h,..$ fixed) such a scale transformation
leads to a transformation of the parameters and the physical
quantities.  Repeated applications of this transformation would
land us on a set of parameters which are functions of $b$.
These are the scale dependent parameters - criticality is best
understood in terms of these scale-dependent parameters.

The scale-dependence of the parameters is best represented by 
infinitesimal scale transformation $b=1+\delta l$
(compare with Ex. III.3).   We may write down a differential equation
for Eq. (\ref{eq:16}) as
\begin{equation}
  \label{eq:51}
 \left (\zeta(t) \ \frac{\partial \ }{\partial t} - y_f  
  \right )\ f =0,
\end{equation}
where $\zeta(t) = {\partial t}/{\partial l}= y_t t$.  This equation  has the
expected solution $f \sim t^{y_f/y_t}$. 

This is a new way of looking at the problem.  Instead of studying a
system for various values of the parameters ( ``coupling constants''), 
we are studying it at various scales to see how it 
behaves in the long-scale limit, after all thermodynamic or
macroscopic behaviour is for a large system.

The equation derived just above, (Eq. (\ref{eq:51}), describes the flow
of the free-energy as the scale is changed, and such equations are
called flow equations.  Any physical quantity will have a flow
equation associated with it. However with only a few relevant
parameters and a few independent exponents, not all flow equations are
necessary.  A fuller picture emerges from a renormalization group
analysis.  Actual flow equations turn out to be slightly more
complicated and the simpler equation of Eq. (\ref{eq:51}) is obtained
under special conditions of ``fixed points''.  The critical behaviour
and the universality classes are  ultimately linked to the fixed points.

\begin{quote}
\begin{exercise} Possibility of a dangerous fixed point?: 
A fixed point by definition is a point that does not change under
rescaling. If a parameter say $u$ has a stable fixed point at $u=u^*$ 
then $u-u^*$ can be taken as an irrelevant variable.  Can this variable be
a ``dangerous irrelevant variable'', or, in other words, can there be
a dangerous fixed point?   
\end{exercise}
\end{quote}

\npg

\section{Models as examples: Gaussian and $\phi^4$}
\label{sec:models-as-examples}

At this point it is helpful to consider a few examples. The Ising
model has been introduced in Ex. III.6.  Here we consider a continuum
version of it.  A naive continuum limit would give a Gaussian model  
for a field variable  $-\infty<\phi<\infty$,
\begin{equation}
  \label{eq:28}
  \frac{H}{k_BT} = \int d^d x \ [ \frac{1}{2}(\nabla \phi)^2 +\frac{1}{2} r
  \phi^2 - h \phi] =  \int \frac{d^d q}{(2\pi)^d} [\frac{1}{2}(q^2 +
  r) \phi_k \phi_{-k}] - h \phi_{k=0}, 
\end{equation}
with a cut-off in real space, i.e.  $|x| >a$ or in momentum space 
$|q|<\Lambda \sim a^{-1}$, where $a$ could be the lattice spacing. We shall 
assume a spherical Brillouin zone. A better continuum limit is the 
$\phi^4$ model,
\begin{equation}
  \label{eq:29}
  \frac{H}{k_BT} =\int d^d x \ [\frac{1}{2}  (\nabla \phi)^2 +
  \frac{1}{2} r  \phi^2 + u \phi^4].
\end{equation}

\begin{quote}
\begin{exercise}
  Starting from the Ising model on a hyper-cubic lattice of
  coordination number $q=2d$, get the continuum form given above.
  Show that $r \sim T-T_m$ with $T_m = q J$.
\end{exercise}
\end{quote}

The reason for considering a continuum limit  is to take
advantage of dimensional analysis.  Both sides of Eqs. \ref{eq:28}
and \ref{eq:29} being dimensionless and the right hand side expressed
as a volume integral helps us in introducing a length based analysis in
a natural way.

That $r=0$ is a singularity is obvious from Eq. (\ref{eq:28}) because
of the instability at $r<0$.  One can derive all the thermodynamic
properties for the Gaussian model and convince oneself that there is a
singularity at $r=0$. We do not go into that.  Here let us
accept  that $r=0$ corresponds to  a critical point.  

\begin{itemize}
\item In general, one would have a term of the type 
$\frac{1}{2}c (\nabla \phi)^2$ in Eqs. \ref{eq:28} and \ref{eq:29}. 
However, if $c$ does not change sign, one may absorb it in the definition of 
$\phi$ and redefine $r$ and $u$.  This has been done in those two equations.
There are problems where $c$ may change sign when external parameters are 
changed (Lifshitz point), and in such cases $c$ needs to be kept explicitly.
It would be necessary to keep $c$ explicitly in case the field variable
cannot be scaled arbitrarily, as e.g. if $\phi$ is like an angular variable.
\end{itemize}

\subsection{Specific heat for the Gaussian model}
\label{sec:Spec-heat-Gauss}

Using Gaussian integrations, one can compute the zero-field specific
heat for the Gaussian model.  The leading term is given by the
integral
\begin{equation}
  \label{eq:30}
  c_{h=0}\approx \xi^{4-d} \frac{Ta_2}{2} \
  \frac{K_d}{(2\pi)^d}\, \int_0^{\Lambda\xi}
  \frac{q^{d-1}}{(1+q^2)^2}\, dq + ... ,
\end{equation}
where $\xi=r^{-1/2}$, $r=a_2(T-T_m)$ and $K_d$ is the surface integral 
of a $d-$dimensional unit sphere. The length-scale exponent is $\nu=1/2$.

\begin{quote}
\begin{exercise}
       Derive Eq. (\ref{eq:30}).\quad Show that $\eta_E = d-2$.
\end{exercise}
\end{quote}

A dimensional analysis gives
\begin{equation}
  \label{eq:31}
  [\phi(x)]=L^{(2-d)/2},\ [r]=L^{-2}, \ [\Lambda]=L^{-1}, {\rm and}\
  [u]=L^{d-4}, [h]=L^{-(2+d)/2}.
\end{equation}
where $L$ is a length scale. This simple dimensional analysis already
identifies a diverging length-scale $\xi=r^{-1/2}$, used in Eq.
(\ref{eq:30}). Note also that the specific heat is a volume integral
of the $\phi^2$-$\phi^2$ correlation function in real space. A
dimensionally correct form is therefore
\begin{equation}
  \label{eq:32}
  c=\xi^{4-d}\ {\cal C}(u\xi^{4-d}, h\xi^{(2+d)/2}, \Lambda\xi).
\end{equation}
We take the Gaussian model first ($u=0$) and no external field, $h=0$.
Then $c=\xi^{4-d} {\cal C}(\Lambda\xi)$ and Eq. \ref{eq:30} is in this
form.  Take the limit $\xi\rightarrow\infty$. The behaviour of the
specific heat  now depends on ${\cal C}(z)$ as
$z\rightarrow\infty$.  If this limit is finite, then the cut-off can
be completely forgotten and $\xi$ plays the important role.  In such a
case the microscopic parameters are not important for the critical
behaviour.  This happens, as we see from the integral in Eq.
(\ref{eq:30}) for $d<4$, and $\alpha=(4-d)/2$, a value that satisfies
hyper-scaling with $\nu=1/2$.  However for $d>4$, the integral
diverges at the upper limit and ${\cal C}(z) \sim z^{d-4}$, leaving us
with a non-divergent specific heat, i.e. $\alpha=0$.  Hyper-scaling
also gets violated. For $h\not= 0$, Gaussian integrations yield
$y_h=(d+2)/2$, consistent with dimensional analysis. The exponents are
\begin{equation}
  \label{eq:52}
\fbox{$y_f=d$ for $d<4$, but $y_f=4$ for $ d>4$} \quad {\rm and}\quad 
\fbox{$\nu=1/2$ and $\eta=0$, for all $d$}.  
\end{equation}

In the above example we chose the specific heat because of its
interesting behaviour for $d<4$ and $d>4$.  Take Susceptibility.
Dimensional analysis gives $\chi \sim r^{-1}$.  Convince yourself, by
using Gaussian integrals, that this is so for all $d$.  However the
finite-size scaling behaviour will be different for $d<4$ and $d>4$.

From the value of $\alpha$ or $y_f$, (see Eq. \ref{eq:52}) and  the
sharpness criterion of Sec. \ref{sec:Role-fluctuations}, we see that
fluctuations can be ignored if $d>4$.  In a sense $d=4$ turns out to
be the upper-critical dimension for this model.

\begin{quote}
\begin{exercise}
  Calculate the average energy of a Gaussian correlated blob. i.e. a
  blob of size $\xi^d$ - do this by calculating the form of the energy
  for the Gaussian model and then putting $r=0$.  Study its behaviour
  for various $d$ and then integrating once with respect to $r$
  estimate the free-energy $f_0$ of a blob (See Sec.
  \ref{sec:faith-thermodynamcis}).  Justify the violation of
  hyper-scaling for $d>4$.
\end{exercise}
\end{quote}

\subsection{Cut-off and anomalous dimensions}
\label{sec:Cut-anom-dimens}

The above warm-up exercise shows that the cutoff, a relic of the
microscopic features of the model, cannot always be ignored even-though
the relevant length-scales at which the phenomenon is taking place is
much much larger than this. Dimensionality is also important.
Dimensional analysis is not expected to 
yield any unique or useful relation if there are multiple scales in
the problem.   This is a very important point and we  elaborate on
this further.   In case the cutoff can be ignored,  the exponents
are the same as  predicted by dimensional analysis.  This we see
directly in the Gaussian model for $d<4$ -  the cut-off 
just didn't matter.

\begin{quote}
\begin{exercise}
  Consider a finite Gaussian model with periodic boundary conditions
  in all directions.  Obtain the finite-size scaling behaviour of the
  specific heat for various $d$.  Note the discrete $k$ sums with no
  $k=0$ mode.
\end{exercise}
\begin{exercise}
  Do the finite-size scaling analysis for the zero-field
  susceptibility of the Gaussian model for various $d$.
\end{exercise}
\end{quote}

If we go back to the $\phi^4$ problem, then we need to treat Eq.
\ref{eq:32}.  Question now is what happens for $\xi\rightarrow\infty$.
If ${\cal C}$ goes to a constant, this cutoff can be safely ignored.
A general situation would be
\begin{equation}
\label{eq:24}
{\cal C}(x,y,z) = z^p \tilde{\cal C}(x z^{p_1}, y z^{p_2}),\quad {\rm
   as}\ z\rightarrow\infty.
\end{equation}
Setting $\Lambda=1$, 
\begin{equation}
  \label{eq:33}
  c \sim \xi^{(4-d)+p} \tilde{\cal C}(u \xi^{(d-4)+p_1}, h \xi^{y_h+p_2}).
\end{equation}
This has the expected scaling form when $\xi$ is replaced by
$|t|^{-\nu}$.  In fact the same analysis can be done for $\xi$ itself,
\begin{equation}
   \label{eq:34}
   \xi= r^{-1/2} \ {\cal X}(u r^{(4-d)/2},...,\Lambda/\sqrt r),
\end{equation}
and, in the $r\rightarrow 0$ limit, the rhs of Eq. \ref{eq:34} might 
(and would) pick up extra powers of $r$ from the $\Lambda$-dependent
argument\footnote{ There will be a shift in the critical temperature
  also. We take $r$ as the deviation from the actual critical
  temperature.}.  This would then change the temperature dependent
exponent of Eq. \ref{eq:33}.

Seen from a thermodynamic point of view, these extra powers of $\xi$
are remarkable because these seem to vitiate dimensional analysis.  An
additional scale is needed, and this comes from the small length-scale
$a\sim1/\Lambda$ whenever there is fluctuations at all length scales.
The ultimate exponent one observes are not what dimensional analysis
based on $\xi$ as the important length-scale would have predicted,
except for Gaussian-type models.  (These dimensional-analysis-based exponents 
are called naive or engineering dimensions.) One needs to understand and 
explain the origin of the ``extra'' contribution like the $p$'s, which are 
to be called {\it anomalous dimensions}.  Once the role of cut-off is 
recognized,  the discrepancy with dimensional analysis (which is infallible 
in any case) vanishes.

Because of the long range nature of the correlations, any local fluctuation
can affect regions away for it.  This is the origin of
scale-invariance or occurrence of power laws (scaling).  However the
occurrence of anomalous dimension is  something more.  In a  Gaussian
type  model, the degrees of freedom can be decomposed into independent
modes ( e.g. by going over to Fourier modes - ``normal coordinates'')
and then we see the emergence of long range correlation in the long
distance limit of $q\rightarrow 0$.  The modes remain independent so
that the fluctuation at one scale determined by $q$ does not affect
the fluctuations at other scales.  For  this reason dimensional
analysis gives correct result.  In contrast for a $\phi^4$ type model,
the  modes for different $q$-values are no longer independent 
(coupled by the $\phi^4$ term) and now a fluctuation at a short
distance scale ( $q$ close to the cut-off) can affect the fluctuations
at longer scales even to $q\rightarrow 0$.  

The mode coupling allows a  small fluctuation at some point in space to 
affect other points and the disturbance seen by any other point is the sum 
over all the paths that connect the two points in question.  For low 
dimensions, this sum can have  fluctuations  and this leads to anomalous 
dimensions.  In high enough dimensions, the availability of a large number 
of paths (phase space volume) helps in averaging out the effects. These two 
cases are separated by the upper critical dimension.

\begin{itemize}
\item Note that $u$ becomes irrelevant for $d>4$ and it could be 
a dangerous irrelevant variable.  The exponents in such cases would depend
on the function also.
\end{itemize}

\subsection{Through correlations}
\label{sec:Through-correlations}

It is reasonable to expect that the scaling variable one sees in a
given problem should be independent of the actual physical quantity
one is looking at.  Therefore in the limit of 
$\Lambda \rightarrow\infty$, anomalous dimension like $p_2$ for $h$ should be
the same for all quantities that depend on $h$.  Let us take the
example of the correlation function, in zero field, at criticality,
\begin{equation}
  \label{eq:35}
  g(x)= \frac{1}{x^{d-2}} {\sf g}(u\, x^{4-d}, \Lambda x),
\end{equation}
suppressing the arguments on the lhs.  Now in the limit 
$\Lambda x\rightarrow\infty$, if 
${\sf g}(...,z) = z^{-\eta} \tilde{\sf g}(....)$,
where the nature of the other arguments are not so crucial for us
right now, we have
\begin{equation}
  \label{eq:53}
  g(x)= \frac{1}{x^{d-2+\eta}} \tilde{\sf g}(...) \quad({\rm setting}\
  \Lambda=1). 
\end{equation}
This $\eta$  changes $y_h=(d+2)/2$ to $y_h=(d+2-\eta)/2$.
Identification can therefore be made: $p_2=-\eta$.

Let us reanalyze the zero field critical correlation function
$g(r,\Lambda)$, where the 
$\Lambda$-dependence has  been made explicit.
Under a rescaling of all lengths $x \rightarrow bx$, we see
$g(x,t=0,\Lambda) = b^{-(d-2)} g(bx,0,\Lambda/b)$.  
The scale factor for the correlation function picks out the exponent
one would expect on 
dimensional analysis.  However if a scale transformation is done that
changes the longer length-scales but not the microscopic ones,
i.e. $\xi \rightarrow b\xi$ but $\Lambda\rightarrow \Lambda$, then
\begin{equation}
  \label{eq:36}
    g(x,0,\Lambda)=b^{-(d-2+\eta)} g(bx,0,\Lambda)
\end{equation}
In analogy with the naive dimension, we now define a scaling dimension
which is the dimension one observes in the long scale limit keeping
the microscopic lengths same.  

Denoting the scaling dimension of $X$ by
${\cal S}_X$ from how $X$-$X$ correlation function scales as in
Eq. (\ref{eq:36}), while naive dimension by $d_X$ ( defined by
dimensional analysis), we have ${\cal S}_{\phi}=-(d-2+\eta)/2$ while
$d_\phi=-(d-2)/2$.  
It is easy to check now that 
\begin{equation}
  \label{eq:9}
  {\cal S}_h + {\cal S}_{\phi}= -d=d_h+d_{\phi}.
\end{equation}
This relation for the every pair of conjugate variables is very
useful.
The scaling behaviour of say Eq. (\ref{eq:33}) 
can then be interpreted as dimensional analysis but {\it with 
scaling dimensions}.  Renormalization group
transformation is a way of doing  a transformation that scales the
long lengths keeping the short ones same, thereby picking the scaling
dimensions.

Generalizing the above discussion, we may define the scaling dimension 
$S_{\psi}$ for any local quantity $\psi({\bf x})$ from the critical
auto-correlation function, i.e. the long distance decay of the
$\psi$-$\psi$ correlation.  The analogue of dimensional analysis would 
tell us that for the critical correlation of any combination of local
variables $\psi_1,\psi_2,...$, we should have 
\begin{equation}
  \label{eq:58}
  \langle\psi_1(0)\psi_2(x_2) ...\psi_n(x_n)\rangle =
  |x_2|^{-S_1-S_2..-S_n} \ {\cal Y}\left (\frac{x_3}{x_2},
    ...,\frac{x_n}{x_2}\right). 
\end{equation}
where $S_i$ is the scaling dimension of $\psi_i$.

\begin{quote}
\begin{exercise}
Why is it that the dimensions add up to $-d$?  When can it be
something different? 
\end{exercise}
\begin{exercise}
Use dimensional analysis to write free-energy $f(r,u,h,\Lambda)$ and
magnetization $m(r,u,h,\Lambda)$ as
\begin{mathletters}
\begin{eqnarray}
  \label{eq:54}
  f&=&\xi^{-d} \tilde f(u \xi^{4-d}, h\xi^{(d+2)/2}, \Lambda \xi),\\
  {\rm and}&&\nonumber\\
\label{eq:54b}
 m&=& \xi^{-(d-2)/2} \tilde M(u \xi^{4-d},
  h\xi^{(d+2)/2}, \Lambda \xi), 
\end{eqnarray}
where $\tilde f, \tilde M$ are unspecified functions. For
$\Lambda\xi\rightarrow\infty$  one expects anomalous dimensions:
\begin{eqnarray}
  \label{eq:55}
f&=&\xi^{-d} \hat f(u \xi^{4-d+p_u}, h\xi^{(d+2)/2+p_h}),\\
  {\rm and} \nonumber\\
 m&=& \xi^{-(d-2)/2+p_m} \hat M(u \xi^{4-d+p_u},
  h\xi^{(d+2)/2+p_h}).  
\end{eqnarray}
Express these new functions in terms of the functions in
Eqs. (\ref{eq:54}) and (\ref{eq:54b}).
Using the relation $m=\partial f/\partial h|_{h=0}$, show that
$p_m=p_h\ \ (=-\eta)$.
\end{mathletters}
\end{exercise}
\end{quote}
\npg

\section{Conclusion}
\label{sec:Conclusion}

We attempted to give an introduction to critical phenomena especially
the idea of scaling, its need and consequences.  The emphasis is on
diverging length-scales.  In case of a diverging length scale, as at a 
critical point, the correlations become long ranged.  Such a point is
characterized by (i) the decay exponents of the correlations ( which could 
have anomalous part like $\eta$ and $\eta_E$ of
Sec. \ref{sec:Its-correlations}), and (ii) the number and nature of
the  relevant variables at that point. 

Any problem that does not have any intrinsic or important length-scale
would behave like a critical system.  This absence of length-scales
leads to power law decays of correlations and also power laws for
other physical quantities.  In low dimensions ( less than the
upper-critical dimension, which could very well be infinite)
fluctuations play an important role near or at the critical point and
show up in the anomalous exponents. In such cases 
(i.e. $d<d_u$), the finite-size effects can be understood in terms of
the finite-size scaling.

Distinction needs to be made between scaling and anomalous dimension.
Scaling is the rule for criticality, originating from a diverging
length scale, while anomalous dimension is seen for fluctuation
dominated cases.

Renormalization group provides the proper framework for analyzing such
phenomena.  We feel it is worthwhile to motivate those ideas behind RG
at the introductory level.  Mean-field theory that ignores
fluctuations could then be placed in the proper perspective as special
(called dangerous) behaviour of certain irrelevant variables.

\acknowledgments

It was a very exciting experience for me to present this set of lectures to 
the participants of the SERC school at MRI Allahabad.  
I thank Abhik Basu, Amit K. Chattopadhaya, Harvey Dobbs, Kavita Jain,
Parongama Sen, Saugata Bhattacharyya  for many comments on the manuscript, 
and  thank Flavio Seno for hospitality at Universit\`a di Padova where the 
final version  was  completed.

\npg


\end{document}